\documentstyle[11pt,newpasp,twoside,epsf]{article}
\begin{document}

\title{NGC 3576 IRS 1 in the Mid Infrared}
\author{C\'assio Leandro D. R. Barbosa}
\affil{IAG-USP, R.do Mat\~ao 1226, 05508-900, S\~ao Paulo, Brazil
cassio@astro.iag.usp.br}
\author{Augusto Damineli}
\affil{IAG-USP, R.do Mat\~ao 1226, 05508-900, S\~ao Paulo, Brazil
damineli@astro.iag.usp.br}
\author{Robert D. Blum}
\affil{Cerro Tololo Interamerican Observatory, Casilla 603, La Serena, Chile
rblum@noao.edu}
\author{Peter S. Conti}
\affil{JILA, University of Colorado, Campus Box 440, Boulder, CO, 80309
pconti@jila.colorado.edu}

\begin{abstract}
We present the results of high-resolution mid-infrared observations of the
source NGC3576 IRS 1. Near diffraction-limited images were taken at the Gemini
South Observatory\footnote{Based on observations obtained at the Gemini
Observatory, which is operated by the Association of Universities for
Research in Astronomy, Inc., under a cooperative agreement with the NSF on
behalf of the Gemini partnership: the National Science Foundation (United
States), the Particle Physics and Astronomy Research Council (United Kingdom),
the National Research Council (Canada), CONICYT (Chile), the Australian
Research Council (Australia), CNPq (Brazil) and CONICET (Argentina).} through
OSCIR's\footnote{This paper is based on observations obtained with the
mid-infrared camera OSCIR, developed by the University of Florida with
support from the National Aeronautics and Space Administration, and operated
jointly by Gemini and the University of Florida Infrared Astrophysics Group.}
filters N (10.8 $\mu$m), 7.9, 9.8, 12.5 and IHW18 (18.2 $\mu$m). The source
IRS1 was resolved into 3 sources for the first time at mid-infrared
wavelengths. For each source we constructed the SED from 1.25 to 18 $\mu$m, as
well the color temperature and the spatial distribution of the dust in the
region. The optical depth of the silicate absorption feature at 9.8 $\mu$m is
presented also.
\end{abstract}

\section{Introduction}

The formation mechanism of massive stars is essentially unknown. This is
mostly due to observational difficulties in finding and establishing an
evolutionary sequence for young stellar objects (YSOs). It is believed that
during the accretion phase, YSOs remain heavily enshrouded in dusty cocoons,
behind hundreds of magnitudes of extinction at visual wavelengths (even at
near-infrared wavelengths for the earliest phases).

A very interesting group of YSOs has been identified in the Galactic giant
HII regions (GHII) M17 (Hanson, Horwarth \& Conti 1997), W43 (Blum, Damineli
\& Conti 1999), W42 (Blum, Conti \& Damineli 2000), W31 (Blum, Damineli \&
Conti 2001) and NGC 3576 (Figueredo et al. 2002). These are luminous YSOs
with excess emission in the K band and have color indexes H-K$>$2. J, H and K
spectra of these objects typically show a featureless continuum. In some
cases the CO 2.3 $\mu$m bandhead is seen in emission or absorption, and in
others FeII and/or H$_{2}$ are seen in emission.

Figueredo et al. (2002) identified two massive YSO candidates in the
position of IRS 1 in NGC 3576. This source was discovered by Lacy, Beck \& Geballe
(1982) and it was recently observed at 10 $\mu$m by Walsh et al. (2001),
but none of their images have enough spatial resolution to resolve the source.

\section{Observations}

Observations were obtained at the Gemini South Observatory 8-m telescope on
December 4th, 2001. OSCIR is based on a Rockwell 128$\times$128 pixel Si:As
BIB detector, with a 0.089 arcsec/pixel plate scale at Gemini. The total
field of view of the camera is 11$\times$11 arcsec and the spatial resolution (FWHM of
the PSF star) is $\sim$0.5 arcsec. The images were taken
through the wide N-band filter (10.8$\mu$m) and IHW18 (18.2 $\mu$m) and the 7.9, 9.8 and 12.5
$\mu$m narrow filters. Flux calibration was performed by taking the flux densities
of the mid-infrared standard star $\alpha$ CMa observed during the night as
part of the baseline calibration program. The uncertainty of these procedures
is estimated to be $\sim$10\%, which is good enough for our purposes.

All images presented have on-source exposure time of 46 seconds, except the
N-band image which has 40 seconds. Background and sky subtraction was done
via the standard chop and nod technique.

\section{Results}

\begin{figure}
\plottwo{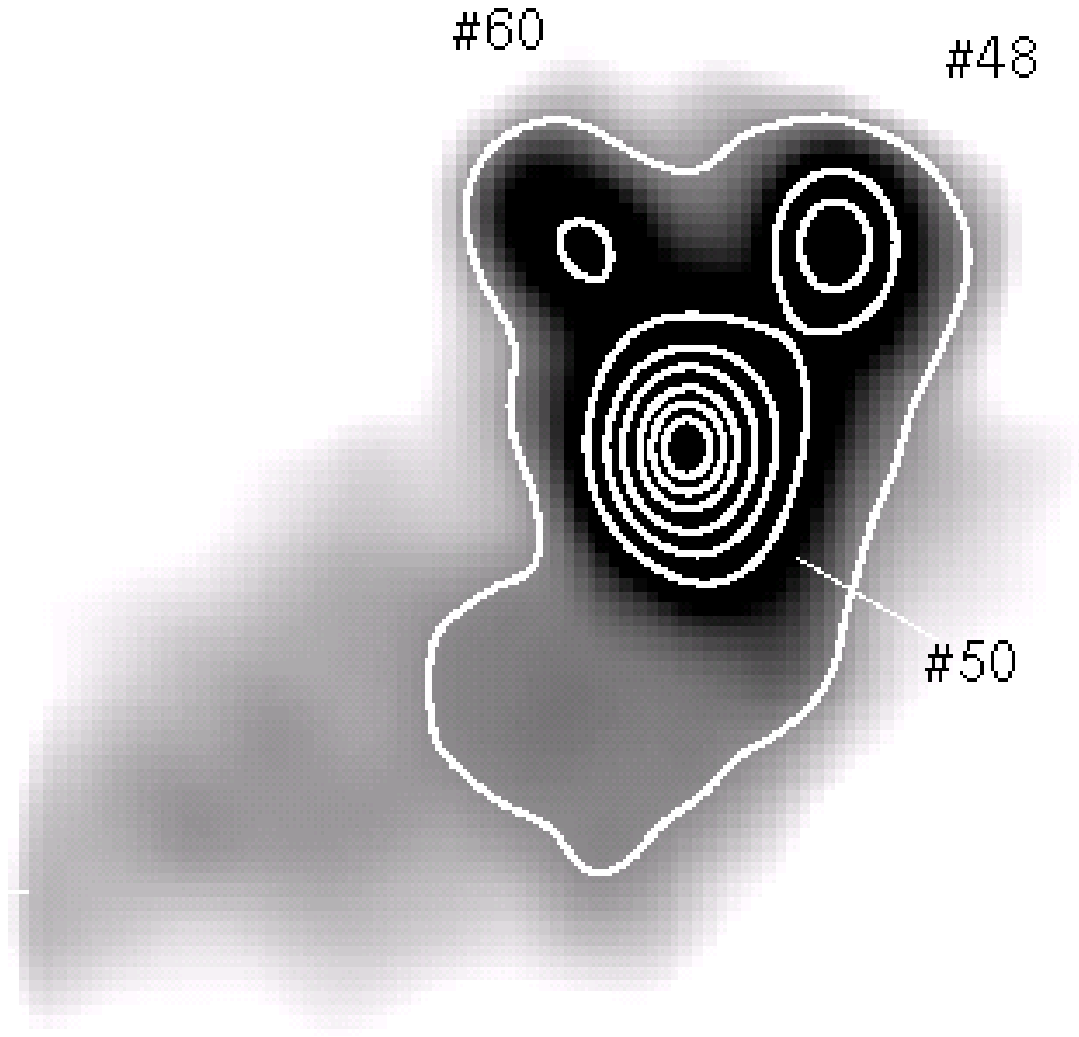}{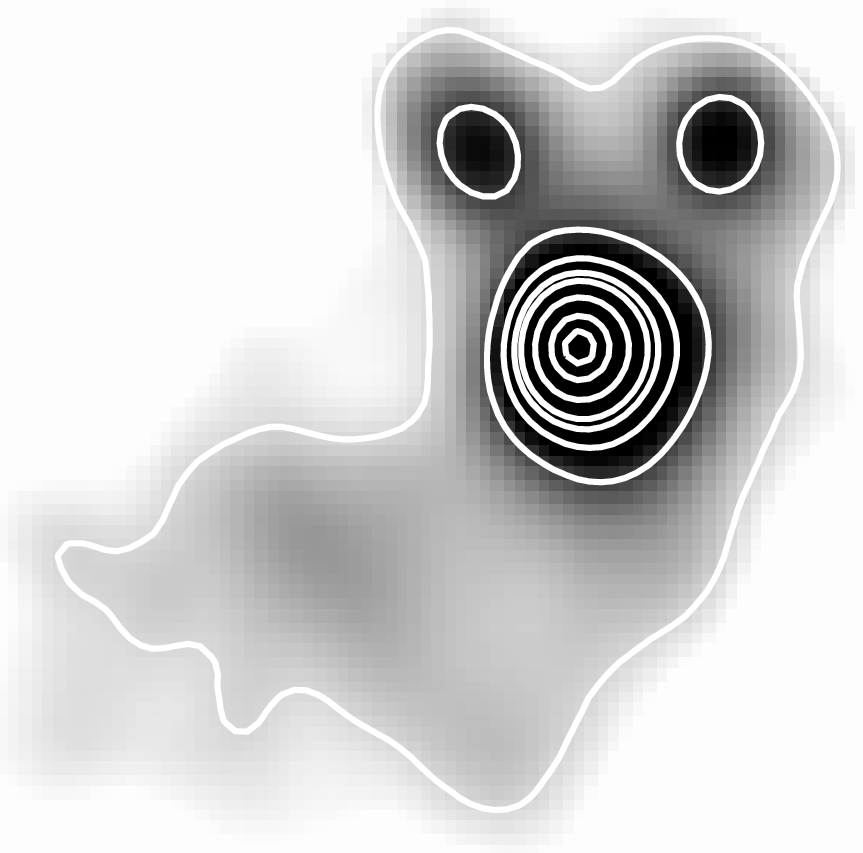}
\caption{{\it Left}:The N-band image of IRS-1. The contours are at 0.01
0.025, 0.05, 0.1, 0.2, ..., 0.5 Jy/ $\Box ^{"}$. {\it Right}: The 18$\mu$m
image of IRS-1. The contours are at 0.044, 0.088, 0.13, 0.17, 0.2, 0.27,
0.35, 0.4 Jy/ $\Box ^{"}$. North is up, East is left for both images. Each
image is 11$\times$11 arcsec.\label{n18}}
\end{figure}

The field of NGC 3576 IRS 1 is shown in the Figure \ref{n18}, for the N and IHW18 bands.
IRS 1 has been resolved into 3 sources embedded in extended emission. All images were smoothed by
convolving each image by the corresponding normalized PSF star fitted from the standard star using
a Fourier transform.

\begin{figure}
\plottwo{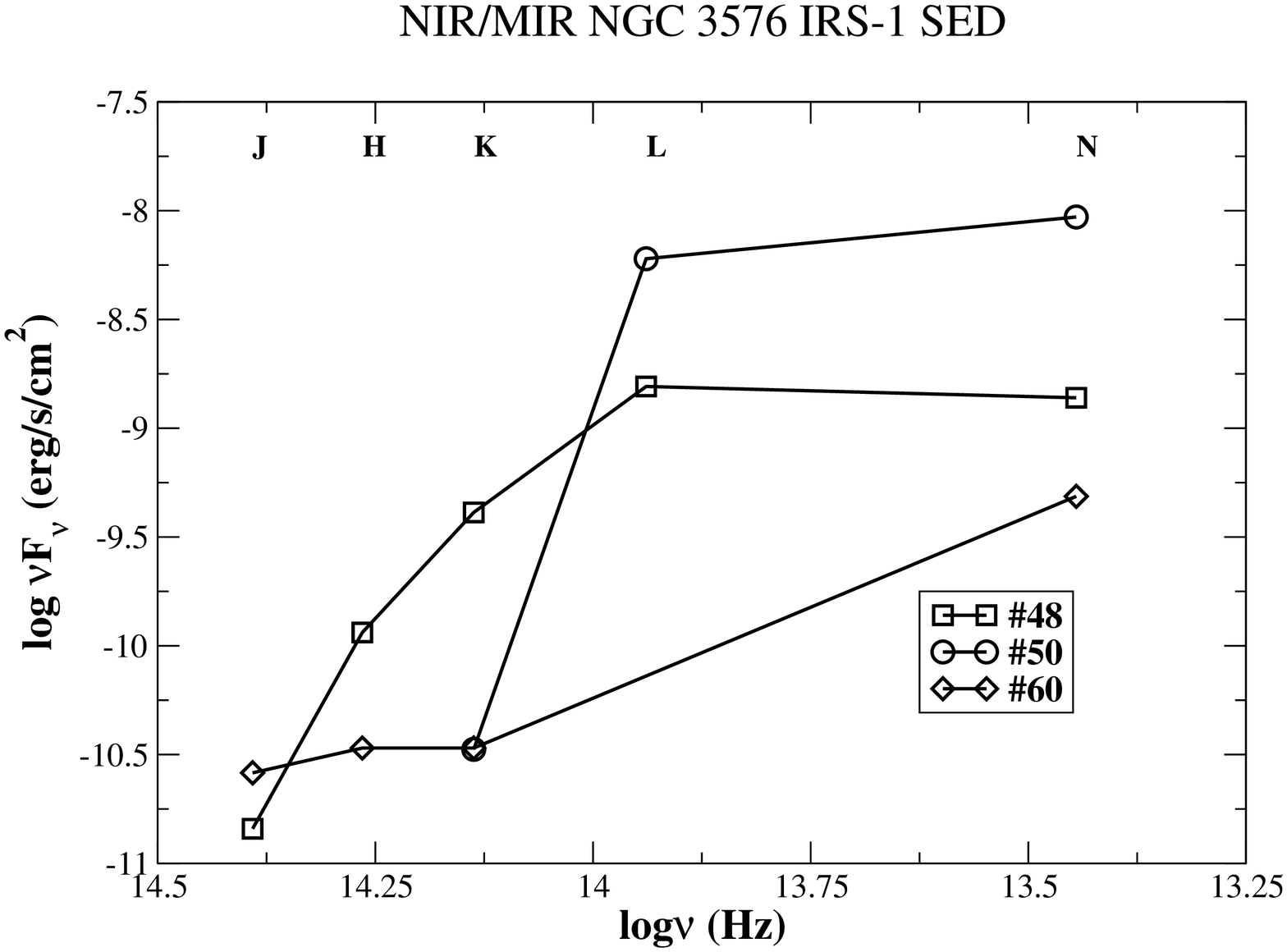}{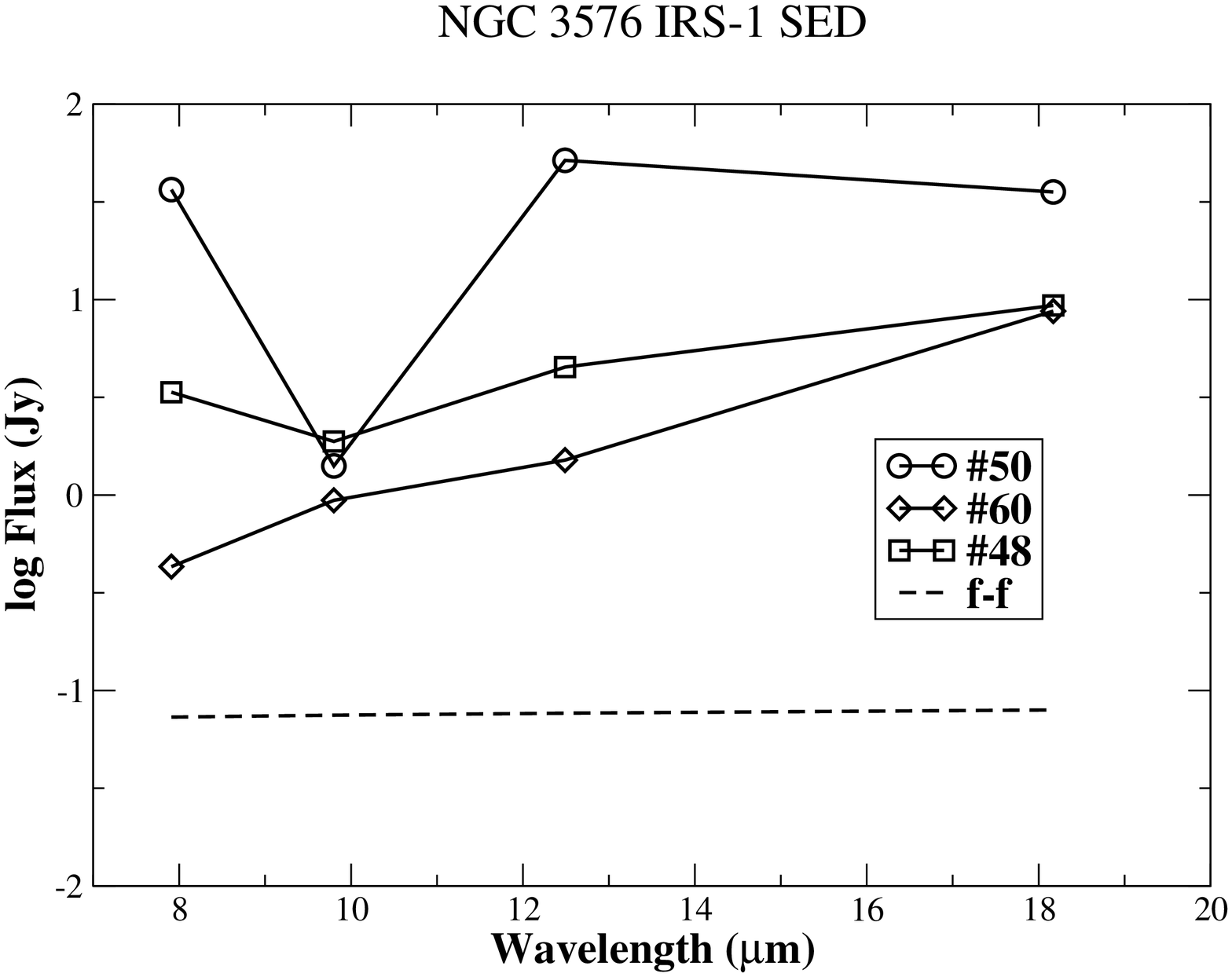}
\caption{{\it Left}:NIR-to-MIR SED for each source identified in the IRS 1
position. {\it Right}: The SED constructed from the fluxes measured at mid
infrared.The dashed line is the projected free-free emission of the gas from
De Pree, Nysewander \& Goss (1999), assuming S$_\nu\propto\nu^{-1}$.
\label{sed}}
\end{figure}

Figure \ref{sed} shows the spectral energy distribution (SED) of each source. The JHK fluxes
are from Figueredo et al. (2002), the L-band flux is from Moneti (1992). The NIR-to-MIR SEDs are
very similar to the results found for a group of less massive AeBe stars by Hillenbrand et al.
(1992). The stars of this group (called {\it group II}) have SEDs with infrared excess and are
supposed to be young stars with intermediate masses (M$\la$ 10 M$_{\sun}$, average spectral type
A5) still accreting. The infrared excess is attributed to a dense dusty accretion disk surrounding
the young stars.

\begin{figure}
\plottwo{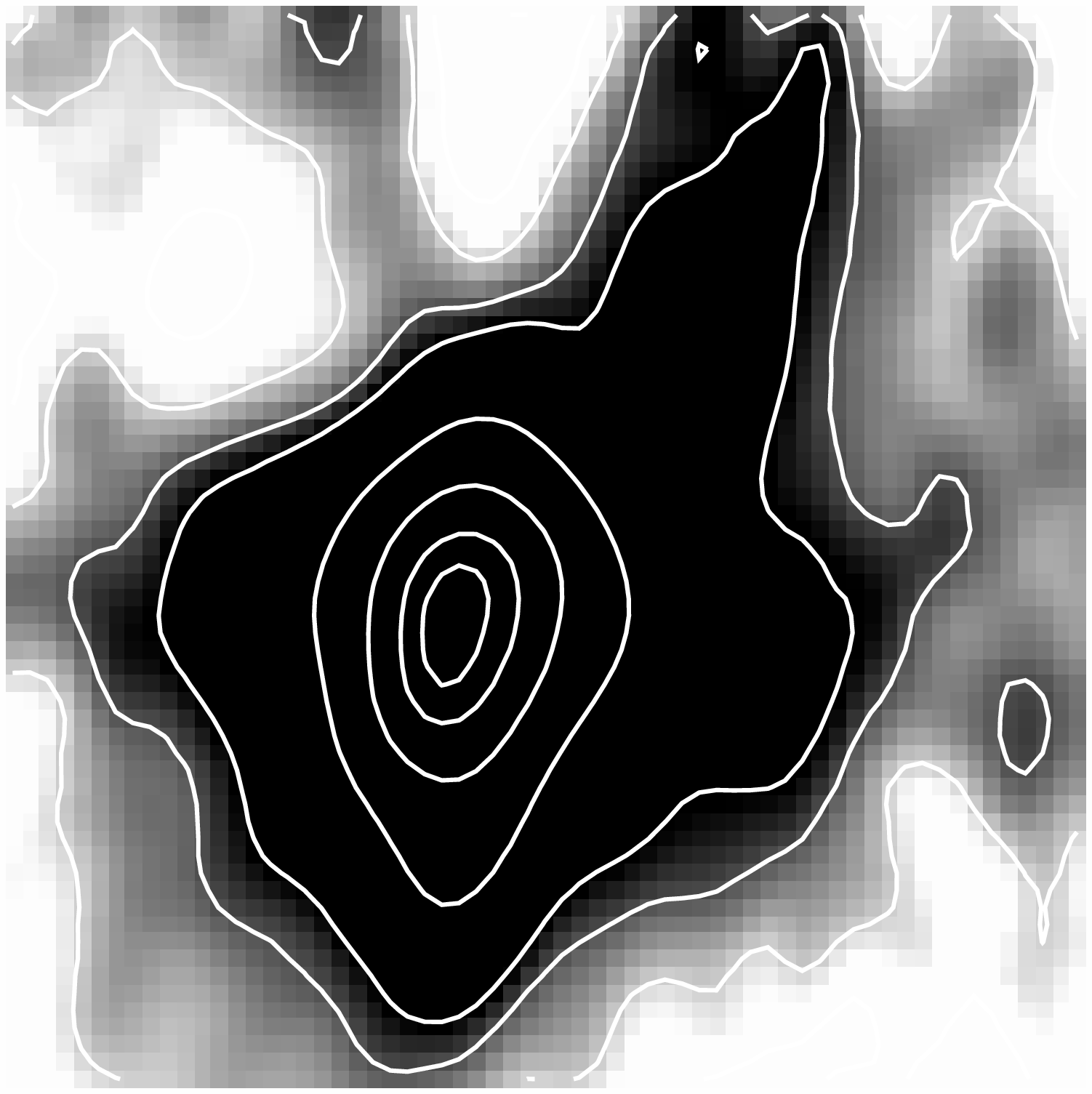}{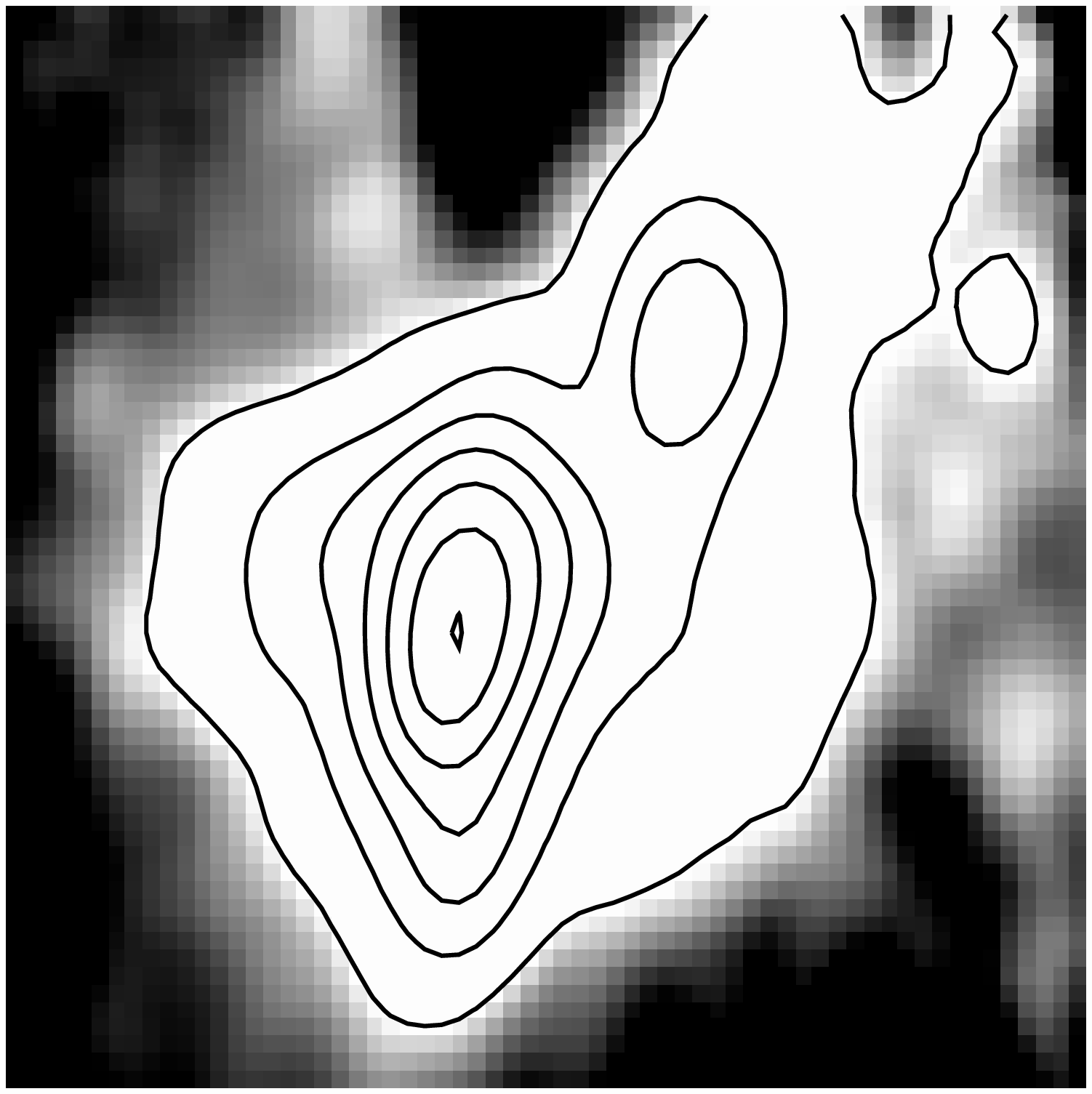}
\caption{{\it Left}: The map of the silicate absorption at IRS-1 position. The
darker is the region, the strongest is the absorption. The contour at
the source \#60 position indicates where the absorption turns into emission.
{\it Right}: The dust color temperature map of IRS-1. The contour levels are
represent temperatures of 160, 180, 200, ..., 280 K. Both images are 5.8$\times$5.8 arcsec.
\label{maps}}
\end{figure}

Figure \ref{maps} shows the spatial distribution of the dust, as well as, its color
temperature map.
The map at the left was obtained dividing the calibrated image taken at 9.8 $\mu$m by the 7.9
$\mu$m image. From this map, we have calculated the optical depth of the silicate absorption
feature and we found $\tau _{9.8}$=3.7. The right panel is the dust color temperature map which
was calculated from the ratio of the 7.9/18 $\mu$m images. From this map we can see the dust
temperature associated with each source: T$\sim$280K (\#50), T$\sim$210K (\#48) and T$<$160K (\#60).

\section{Summary}

We presented high-resolution mid-infrared images of NGC 3576 IRS 1, that has been resolved
into 3 sources for the first time at mid-infrared wavelengths. The SEDs of each source were
constructed from 1.25 to 18 $\mu$m with data taken from literature. Each SED shows increasing fluxes
toward to longer wavelengths. Hillenbrand et al. (1992) found similar SEDs for a group of less
massive young stars in process of accreting mass via an accretion disk embedded in a dusty cocoon.
This similarity suggests the same interpretation for our results. Finally we presented maps of dust
distribution, temperature and optical depth as well.\\
\\
\\
\\
We thank Chris De Pree for kindly providing the FITS file of the 3.4 cm map of NGC 3576.\\
CLB and AD acknowledge the financial support from PROAP, PRONEX and FAPESP.\\
PC appreciates continuing support from the National Science Foundation.

\end{document}